# Holographic time-resolved particle tracking by means of three-dimensional volumetric deconvolution

**Tatiana Latychevskaia\* and Hans-Werner Fink**
*Physics Institute, University of Zurich, Winterthurerstrasse 190, 8057 Zurich, Switzerland*
*\*tatiana@physik.uzh.ch*

**Abstract:** Holographic particle image velocimetry allows tracking particle trajectories in time and space by means of holography. However, the drawback of the technique is that in the three-dimensional particle distribution reconstructed from a hologram, the individual particles can hardly be resolved due to the superimposed out-of-focus signal from neighboring particles. We demonstrate here a three-dimensional volumetric deconvolution applied to the reconstructed wavefront which results in resolving all particles simultaneously in three-dimensions. Moreover, we apply the three-dimensional volumetric deconvolution to reconstructions of a time-dependent sequence of holograms of an ensemble of polystyrene spheres moving in water. From each hologram we simultaneously resolve all particles in the ensemble in three dimensions and from the sequence of holograms we obtain the time-resolved trajectories of individual polystyrene spheres.

## 1. Introduction

Holographic particle image velocimetry (HPIV) is an established technique for tracking particle trajectories in time and space by means of optical holography [1-4]. Off-axis [5-9], inline [10-31] and hybrid schemes [12, 32] of experimental arrangements have been proposed allowing a three-dimensional reconstruction of the wave scattered by the particle field. One of the critical issues in HPIV is assigning locations of individual particles from the reconstructed scattered wave. While the lateral position of a particle in the plane orthogonal to the optical axis can relatively easy be determined, the precise position of the particle along the optical axis (z-axis) is difficult to allocate due to the extended depth-of-focus signal along z-direction. In an optical system, the image of an ideal point source appears as an Airy disk with the central maximum remaining pronounced over an extended defocusing distance along the z-axis. In fact, 80% of the maximal intensity of the central peak is still apparent at the defocus distance δ [2, 32-33]:

$$\delta = \frac{\lambda}{2\Omega^2} \quad (1)$$

where Ω is the effective angular aperture (half-angle) of the hologram and λ being the wavelength. The depth of focus of an ideal point source thus amounts to 2δ. For realistic size-limited particles, the depth of focus is even larger and can reach 40 times of the particle size [15, 20]. The axial position of the reconstructed particle can roughly be assigned to the maximum of the intensity of the reconstructed wavefront [13-14, 19, 28] or by a threshold of the reconstructed intensity [17-18, 34]. Several techniques have been proposed for trying to find the exact position of a particle in axial direction. It has been found that the amplitude of the reconstructed complex-valued wave exhibits oscillations in axial directions with the minimum at the exact particle position [35]. Pan *et al* demonstrated that the particle position can also be found from the minimum of the variance of the reconstructed imaginary part of the scattered wave [15]. Fournier et al. proposed to implement a window-like function into the reconstruction algorithm which reduces the number of oscillations in the reconstructed amplitude profile and thus allows finding the particle position more precisely [16]. Also, the entropy method [23] and Hough transform [24] were proposed for finding the axial positions of particles. Alternatively, the position of the particle and its shape can be recovered by fitting the holographic image with the interference pattern predicted by Mie theory [34, 36]. In practice, these methods are time-consuming and require the analysis of the wave front distribution of each particle. Recently, we demonstrated that all the positions of particles distributed in three dimensions can be retrieved at once from a holographic reconstruction by three-dimensional volumetric deconvolution with the point-spread function of the optical system [25]. Here we show how this three-dimensional deconvolution can be applied to obtain the positions of an ensemble of moving particles.

## 2. Experimental

The experimental setup for a HPIV measurement is shown in Fig. 1. A drop of aqueous solution of 10 μm diameter polystyrene spheres is immersed into a cuvette filled with distilled water and the moving spheres are imaged by inline holography. The quartz glass cuvette, placed right adjacent to the microscope objective, exhibits an interior width of 10 mm and a wall thickness of 1.25 mm. The hologram size projected by the microscope objective onto the screen amounts to roughly 25 mm in diameter and the field of view is about 625 μm × 625 μm. The camera allows capturing digital 10 bit images of 1000 × 1000 pixels corresponding to about 16 pixels for each 10 μm diameter polystyrene sphere. The concentration of the spheres in the drop was empirically selected such that the presence of several flowing spheres could be observed at any time in the scene. For obtaining time-resolved trajectories of the moving spheres, a sequence of 500 frames was recorded with an exposure time of 20 ms per frame and a time interval between two successive images of 110 ms. For the particle-tracking analysis presented here 30 frames were selected.

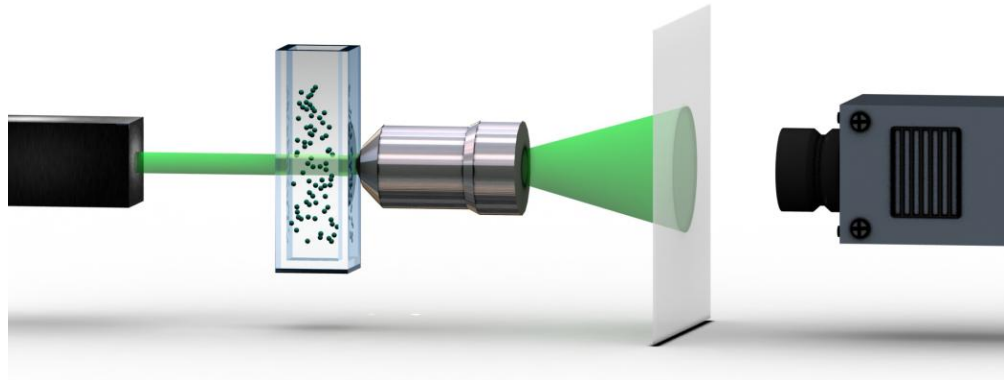

Fig. 1. Experimental scheme for recording optical inline holograms. Laser light of 532 nm wavelength passes through a cuvette filled with distilled water. A drop of aqueous solution of 10 μm diameter polystyrene spheres is immersed into a cuvette and the moving spheres are imaged by means of a microscope objective (magnification = 40, N.A. = 0.65) which projects the hologram of the moving spheres onto a screen made up of a translucent Mylar-like material.

### 3. Hologram normalization

For hologram normalization [37-38] a background image is needed. It is provided by summing up all the recorded holographic frames which smears out the signal from the moving spheres but maintains pronounced signals from static object such as scratches on the cuvette surface, see Fig. 2. Normalized holograms were subsequently obtained by dividing the experimental holograms with the background image.

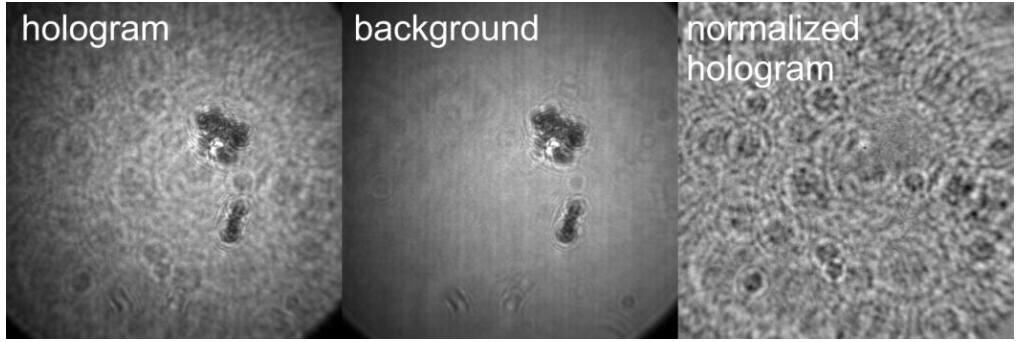

Fig. 2. An individual hologram out of a sequence of experimental holograms is shown at left, the background image in the middle and the normalized hologram at right. The objects apparent in the background are due to scratches on the cuvette surface.

### 4. Reconstruction of holograms

The reconstruction of the normalized holograms is based on applying the Kirchhoff scalar diffraction theory with the Fresnel-Sommerfeld solution. The Huygens-Fresnel principle, as predicted by the Fresnel-Sommerfeld solution, can be expressed as follows [39]:

$$U(\mathrm{P}_0) = \frac{1}{i\lambda} \iint_\Sigma U(\mathrm{P}_1) \frac{\exp ikr}{r} \cos \vartheta \, ds, \qquad (2)$$

where $\lambda$ is the wavelength, $\vec{r}$ denotes the vector from point $\mathrm{P}_1$ to point $\mathrm{P}_0$, and $\vec{n}$ denotes the vector normal to the hologram surface, $\theta$ the angle between vectors $\vec{n}$ and $\vec{r}$, $U(\mathrm{P}_0)$ the reconstructed complex-valued field, $U(\mathrm{P}_1)$ the complex-valued field at the hologram, and the integration is performed over the hologram surface. In the experiments presented here, the Fresnel approximation $z^3 \gg \frac{\pi}{4\lambda}\left[(x-X)^2+(y-Y)^2\right]^2_{\max}$ is not fulfilled; here $z$ is the distance between sample and detector, and $(x,y)$ and $(X,Y)$ are the coordinates in the object and detector plane respectively. Since the Fresnel approximation is not fulfilled, we apply the angular spectrum theory [39] for a plane wave propagating short distances. The complex-valued wave scattered by the object $U_{\text{Object}}(x,y,z)$ is reconstructed from the hologram by back-propagation of the optical field from the hologram plane $(X,Y)$ to the object plane $(x,y)$ [39]:

$$U_{\text{Object}}(x,y,z) = \mathrm{FT}^{-1}\left[\mathrm{FT}\{H(X,Y)\}\exp\left(-\frac{2\pi i z}{\lambda}\sqrt{1-(\lambda f_x)^2-(\lambda f_y)^2}\right)\right], \qquad (3)$$

where FT and $\mathrm{FT}^{-1}$ are the Fourier transform and inverse Fourier transform, respectively, and $(f_x, f_y)$ are the spatial frequencies.

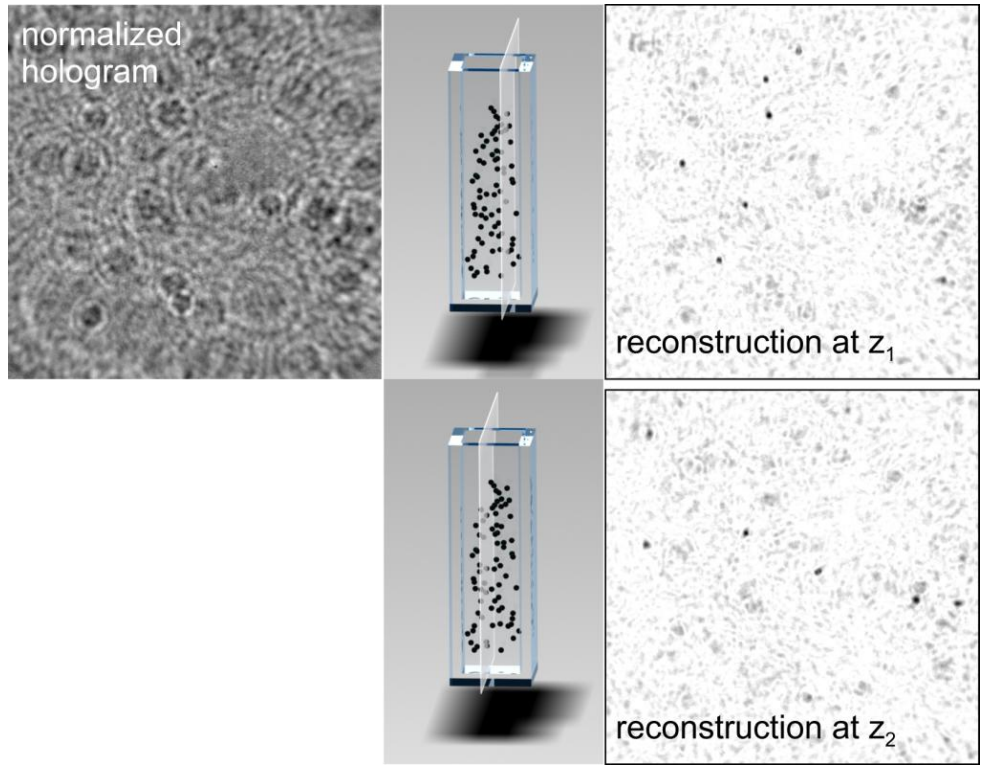

Fig. 3. Amplitude of the object wave distribution reconstructed from one hologram out of the sequence of holograms. Reconstructions at z = 2.66 mm and 4.22 mm from the hologram plane are shown.

Figure 3 shows the amplitude of the object wave reconstructed from one hologram out of the sequence of holograms. It shows the distribution of spheres in the imaged volume at a certain time but at different planes within the sampled volume. The reconstructions displayed in Fig. 3 demonstrate the challenge when dealing with an ensemble of particles flowing in a solvent. There is significant background noise that we associate with diffuse scattering in the solvent and possible interference between internally reflected light from the cuvette surfaces. It manifests itself also in the very noisy appearance of the normalized holograms, as shown in Fig. 2 and Fig. 3, which does not allow fitting the hologram of each sphere using Mie theory.

## 5. Three-dimensional volumetric deconvolution

The reconstruction of a hologram results in the three-dimensional distribution of the scattered wave. The positions of individual scatterers can be obtained by applying a three-dimensional deconvolution of the reconstructed wave front with the point-spread function (PSF) of the optical system The latter is a complex-valued three-dimensional wave scattered by a point-like object and imaged by the optical system [25]. In [25] we proposed two deconvolution methods: instant and iterative deconvolution and demonstrated their application to experimental holograms of microspheres distributed three-dimensionally. As a result, we were able to obtain the three-dimensional distribution of particle positions at increased resolution in both axial and lateral directions. The algorithm of instant three-dimensional volumetric deconvolution was applied by Dixon *et al* [27] for tracking of a few objects in the scene. In the work presented here we apply the method of instant three-dimensional deconvolution of the *intensity* of the reconstructed wavefront with the *intensity* of the PSF of the optical system. Each hologram in the experimental sequence was resampled with 200 × 200 pixels and reconstructed at a series of distances from the hologram ranging from 1 to 7 mm with a step width of 0.3 mm to obtain a total of 200 reconstructions. Thus, from each hologram we obtain the reconstructed three-dimensional volumetric complex-valued field sampled with 200 × 200 × 200 pixels. The number of pixel is limited by the performance of the MATLAB fftn routine employed for three-dimensional deconvolution. The deconvolution was performed on 64-bit operating system with 16 Gb RAM.

### a. PSF

The three-dimensional volumetric deconvolution method does not require any calibration experiments and the PSF of the system is obtained by reconstructing either a simulated hologram of a point scatterer or an experimental hologram of a single scatterer (a cutout of an experimental hologram where just a single scatterer is observed) [25]. In the experiments presented here, the PSF was obtained as the three-dimensional complex-valued wavefront reconstructed from a simulated hologram of a point scatterer, see Fig. 4. The hologram for obtaining the PSF is simulated as follows: a point scatterer is placed at z = 4 mm from the detector plane and its hologram is calculated with the following formula:

$$H_{PSF}(X,Y) = \left| \mathrm{FT}^{-1}\left[ \mathrm{FT}\{1-\delta(x,y)\} \exp\left( \frac{2\pi i z}{\lambda}\sqrt{1-(\lambda f_x)^2-(\lambda f_y)^2} \right) \right] \right|^2, \quad (4)$$

where $\delta(x,y)$ is the point-scatterer described by a delta-function and assumed to be of the size of one pixel for the numerical simulation. The parameters of the simulated hologram are identical to those of the experiment: a wavelength $\lambda$ = 532 nm and an imaged area of 625 μm × 625 μm. The simulated hologram is reconstructed by back propagating the wavefront calculated by Eq. (3):

$$U_{PSF}(x,y,z) = \mathrm{FT}^{-1}\left[ \mathrm{FT}\{H_{PSF}(X,Y)\} \exp\left( -\frac{2\pi i z}{\lambda}\sqrt{1-(\lambda f_x)^2-(\lambda f_y)^2} \right) \right], \quad (5)$$

and the reconstructed complex-valued wavefront $U_{PSF}(x,y,z)$ constitutes the PSF.

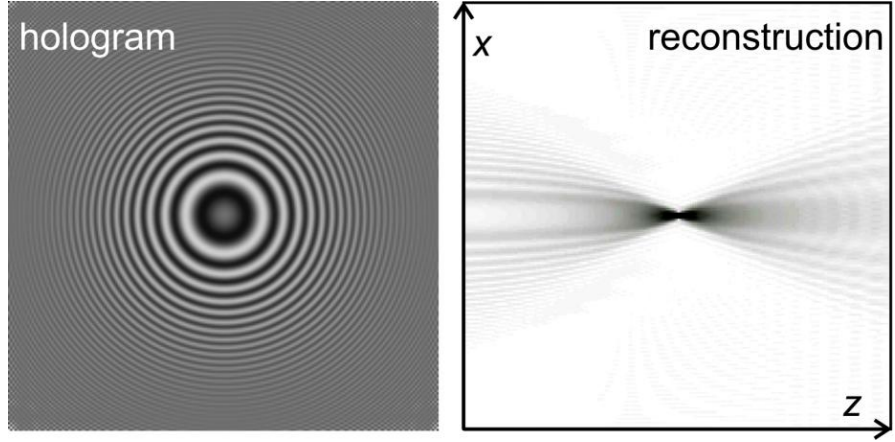

Fig.4. Simulated hologram of a point scatterer and the distribution of the reconstructed amplitude in the *XZ* plane.

### b. Results of three-dimensional volumetric deconvolution

Three-dimensional volumetric deconvolution is calculated by applying the following formula to the three-dimensional distributions [25]:

$$o(x,y,z) = A_O(x,y,z) \text{FT}^{-1}\left( A_F(X,Y,Z) \frac{\text{FT}\left[I_{\text{Object}}(x,y,z)\right]}{\text{FT}\left[I_{\text{PSF}}(x,y,z)\right]} \right) \approx$$

$$\approx A_O(x,y,z) \text{FT}^{-1}\left( A_F(X,Y,Z) \frac{\text{FT}\left[I_{\text{Object}}(x,y,z)\right]\left\{\text{FT}\left[I_{\text{PSF}}(x,y,z)\right]\right\}^*}{\left|\text{FT}\left[I_{\text{PSF}}(x,y,z)\right]\right|^2 + \beta} \right) \quad (6)$$

where $I_{\text{Object}}(x,y,z) = |U_{\text{Object}}(x,y,z)|^2$, $I_{\text{PSF}}(x,y,z) = |U_{\text{PSF}}(x,y,z)|^2$, and $\beta$ is a small constant added to avoid division by zero. In this three-dimensional deconvolution, a numerical apodization window function $A_F(X,Y,Z)$ is applied in the Fourier domain:

$$A_F(X,Y,Z) = 1 \text{ if } \sqrt{X^2 + Y^2 + Z^2} \leq N_F,$$
$$0 \text{ if } \sqrt{X^2 + Y^2 + Z^2} > N_F, \quad (7)$$

where $N_F$ is the cut-off frequency in pixels; we used $N_F = 70$. $A_F(X,Y,Z)$ cuts off higher order frequencies that are mostly due to noise. The second numerical filter $A_O(x,y,z)$ is applied in the object domain:

$$A_O(x,y,z) = 1 \text{ if } \sqrt{x^2 + y^2 + z^2} \leq N_O,$$
$$0 \text{ if } \sqrt{x^2 + y^2 + z^2} > N_O, \quad (8)$$

whereby $N_O$ denotes the cut-off limit in pixels and we used $N_O = 95$. This filter allows avoiding accumulation of artefacts at the edges of the three-dimensional reconstruction volume. This three-dimensional volumetric deconvolution results in the three-dimensional distribution of the positions of the individual scatterers. In Fig. 5, the reconstructions of three consecutive holograms before and after deconvolution are shown.

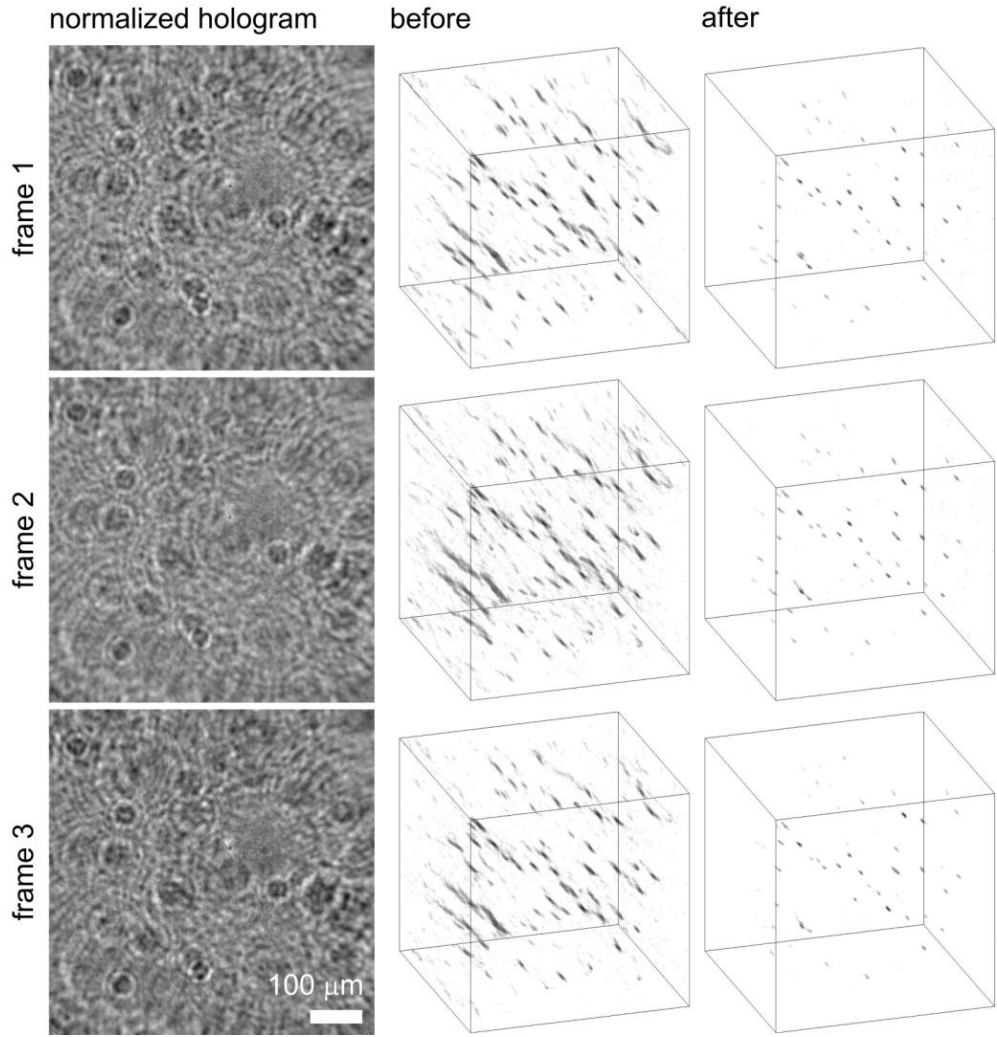

Fig. 5. Three successive holograms out of the sequence and their three-dimensional reconstruction before and after applying the three-dimensional volumetric deconvolution are shown. The area of the reconstructed volume amounts to 625 μm × 626 μm × 6000 μm.

In Fig. 6, an intensity profile of a single scatterer before and after three-dimensional deconvolution is displayed. Obviously, the resolution in both, axial in lateral directions is greatly improved. As evident from Fig. 6(b), the lateral positions of scatterers before deconvolution are represented as blurred spots. After deconvolution they appear sharp and limited in size to just 2 or 3 pixels, as shown in Fig. 6(c). Another advantage of the three-dimensional deconvolution with the three-dimensional PSF is that it allows distinguishing between the intensity maximum caused by the particle and its caustics. Caustics are maxima of intensity formed by the superposition of the waves scattered by neighboring particles, they do not possess the out-of-focus distribution of a particle, see Fig. 4. The deconvoluted distribution exhibits a maximum only when the signal scattered by a particle perfectly matches the PSF being the signal from a perfect point scatterer including its out-of focus distribution, see Fig. 4. The caustics will thus not result in maxima in the three-dimensional deconvolution. As evident from Fig 6, some of the intensity maxima are missing after deconvolution; they are attributed to caustics formed by the superposition of waves scattered by the neighboring particles, as for example indicated by the green arrow in Fig. 6(b) and 6(c). The three-dimensional volumetric deconvolution is thus a much more accurate technique than other methods for determining particle positions. The best condition for a particle to be resolved by three-dimensional deconvolution is when the particle is not too close to an edge of the reconstruction volume since part of its three-dimensional signal is then cut off. In this case the match between the three-dimensional PSF and three-dimensional particle reconstruction is only partial and the deconvolution might miss it.

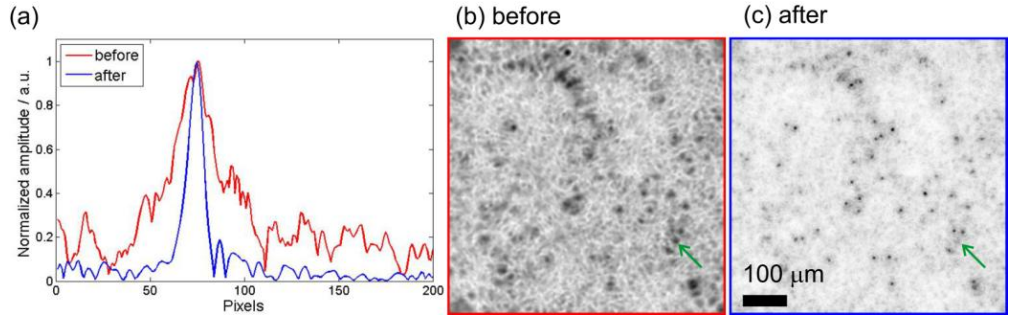

Fig. 6. Reconstructions before and after performing a three-dimensional volumetric deconvolution. (a) Profiles of the reconstructed intensity distribution of an individual scatterer along the axial direction (z-axis). (b) Reconstructed intensity distributions summed up along axial direction before a three-dimensional deconvolution and (c) after three-dimensional deconvolution.

## 6. Time-resolved particle tracking

A time-dependent sequence of holograms allows observing the evolution in the three-dimensional distribution of particles. Such time-resolved holographic particle tracking has various applications. When only a few particles are present in the scene, their movement can easily be observed by the naked eye and matching of each particle position in different time frames can be done manually. In this manner, for example, evaporation process of droplets in jets [31] and velocity profiles in a tube showing parabolic distribution [19] have been studied. When a larger number of scatterers is flowing in a solvent, as for example when studying flow dynamics, the velocity distribution is typically obtained by the cross-correlation between particle positions [8]. However, the method of cross-correlation requires the position of each particle as an input, and those positions are difficult to find when the number of particles is large. Conventionally, positions of individual particles are extracted and stored as coordinates (x,y,z). Then, a three-dimensional volume is created where pixels corresponding to particle positions are marked. This creates a three-dimensional representation of the particle distribution. In our method, after a three-dimensional deconvolution, single particles are clearly resolved and their trajectories can be followed individually, see Fig. 5. As an example, we have selected four spheres which are close to each other but exhibit different sedimentation velocities, as illustrated in Fig. 7.

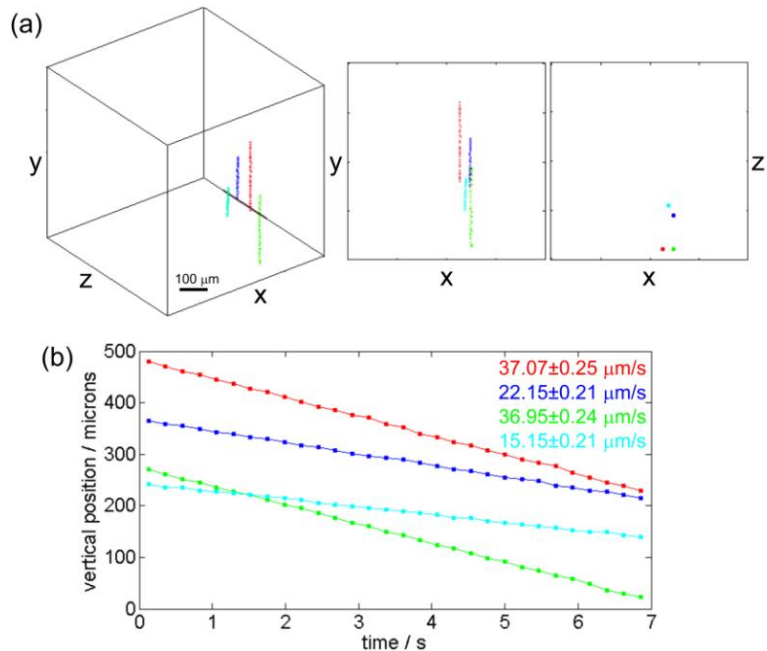

Fig. 7. Trajectories of four individual microspheres. (a) Three-dimensional and projected views. The area of the reconstructed volume amounts to 625 μm × 626 μm × 6000 μm. (b) Vertical position of the particles as a function of time. The sedimentation velocity of each particle is estimated from the linear fitting. The pairs of closely spaced particles, red–green (34 μm distance in xz-plane) and blue–cyan (300 μm distance in xz-plane) exhibit similar velocities.

## 7. Conclusions

We have demonstrated that the three-dimensional volumetric deconvolution can be successfully applied to simultaneously resolve individual positions within an ensemble of particles moving in a solvent. Good reconstructions can be achieved despite the interference between numerous particles and the background due to the solvent. The three-dimensional volumetric deconvolution allows separating between the signal originating from a particle and the caustics. After deconvolution, particle positions are clearly resolved and their tracking allows following their time-resolved trajectories.

## Acknowledgment

The Swiss National Science Foundation and the University of Zurich are gratefully acknowledged for their financial support. We would also like to thank Jessica Britschgi and Alina Horwege for their assistance with the experimental setup.